# Beyond Efficiency and Convenience. Using Post-growth Values as a Nucleus to Transform Design Education and Society

Matthias Laschke

University of Siegen, Interaction Design for Sustainability and Transformation, matthias.laschke@uni-siegen.de

Lenneke Kuijer

Eindhoven University of Technology, Department of Industrial Design, s.c.kuijer@tue.nl

In this position paper we present Municipan, an artefact resulting from a post-growth design experiment, applied in a student design project. In contrast to mainstream human-centered design directed at efficiency and convenience, which we argue leads to deskilling, dependency, and the progression of the climate crisis, we challenged students to envision an opposite user that is willing to invest time and effort and learn new skills. While Municipan is not a direct step towards a postgrowth society, integrating the way it was created in design education can act as a nucleus, bringing forth design professionals inclined to create technologies with potential to gradually transform society towards postgrowth living. Bringing in examples from our own research, we illustrate that designs created in this mindset, such as heating systems that train cold resistance, or navigation systems that train orientation have potential to reskill users, reduce technological dependency and steer consumption within planetary limits.

## 1 DOES TECHNOLOGY MAKE PEOPLE'S LIVES BETTER?

A prevailing narrative within design and HCI is that innovative technology is key for prosperity and wellbeing. To achieve this, designers are expected to follow a human-centered approach. In this approach, people's goals and needs are at the center of all design efforts, and resulting designs are assumed to make their lives better. In the following, we take a closer look at this "better" and discuss how the current conception designers have of people's needs and goals creates an unsustainable form of "better".

Looking at everyday life, it becomes apparent that almost every daily practice is performed with and through technology. A status that Peter-Paul Verbeek, a philosopher of technology, refers to as a life that is "unthinkable without sophisticated technology" (Verbeek, 2011). Technology allows people to do things that they could not do without them, do them faster, or with less effort. For instance, vacuuming requires a vacuum cleaner. Without it, people would probably sweep (although a broom is a technology too) or collect dust with their bare hands. Thus, technology takes over (part of) a skill (i.e., to clean). Latour refers to this mechanism as 'delegation' (Latour, 1992). Compared to sweeping or manual cleaning, vacuum cleaning is more efficient and convenient. Moreover, the delegation of skills to technology proves to be a lucrative business model, exemplified by the development of robot vacuum cleaners (i.e., a further step in the development of the vacuum cleaner). These devices eliminate the need for cleaning (both efficient and convenient), albeit at a significantly higher cost. Unfortunately, as (Tonkinwise, 2018) points out, once people "tasted" the efficiency and convenience of new technological options, they rarely revert to supposedly less efficient or inconvenient options.

Throughout technological innovation, spanning from basic tools like brooms to sophisticated devices like robot vacuum cleaners, there has been a persistent drive towards achieving greater efficiency and convenience. This pattern of technological innovation, marked by the continuous introduction of more advanced and efficient versions of technology across various aspects of life, makes everyday life increasingly resource intensive. Moreover, it perpetuates a growing inclination among people to rely on technology as a means of enhancing efficiency and convenience, forming part of a neoliberal 'good life' narrative – the lead character of which Dahlgren et al. (2021) refer to as *techno-hedonist persona* – living a clean, streamlined, smooth and regular life. But can such a life be called *better*?

Research is increasingly indicating the contrary. Besides increasing burdens on the planet, **delegating capabilities to technology deskills people**. For instance, using a turn-by-turn navigation system will probably result in people no longer being able to find their way around a city. Western philosophers have long criticized this point of deskilling in the adoption of technology: writing lowers memory power, television lowers imagination, smartphones screens lower sociality (Albert Borgmann, 1984; Harry Braverman, 1998).

This loss of capabilities **makes people dependent on technology**. Not only in a practical sense, as people require all these technologies to perform a practice, but also in an emotional sense, as their success and wellbeing depend on being able to perform practices in an (ever more) efficient and convenient way. We consider such an approach to a better life to be rather "unsustainably better". The logic of technological innovation, which seeks to increase efficiency and convenience through new versions of technology (Pirgmaier, 2020; Strand et al., 2018) **drives consumerism and thus climate change**.

In this position paper, we reflect on possible design responses that adopt a different conception of what users want, starting with Municipan.

## 2 MUNICIPAN. A NUCLEUS TO TRANSFORM DESIGN EDUCATION AND PRACTICE

Municipan was created by a group of students as part of the first-year master course *Constructive Design Research*. The course covers design research methods, which students then apply to research topics suggested by their supervisor. The group was tasked to imagine a persona contrasting with the techno-hedonist persona by (Dahlgren et al., 2021) and considering the implications for interaction design practices. The students addressed this challenge through a first person, reflective, making oriented process focused on cooking practices.

The students began by imagining an alternative present in which the microwave, an ultimate convenience technology, had been rejected at the time of its emergence. This thought experiment resulted in a scenario encompassing community farming and communal cooking (Figure 1a). Municipan (Figure 1b) is designed as a part of this alternative reality.

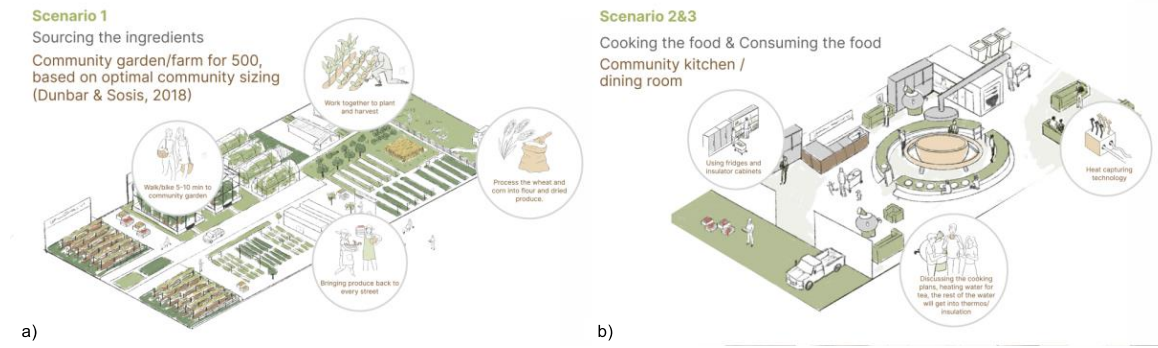

Figure 1: a. An alternative communal living system where food is grown locally (Dunbar & Sosis, 2018). b. Various facilities and Municipan, a pan in which only several portions of a dish can be cooked.

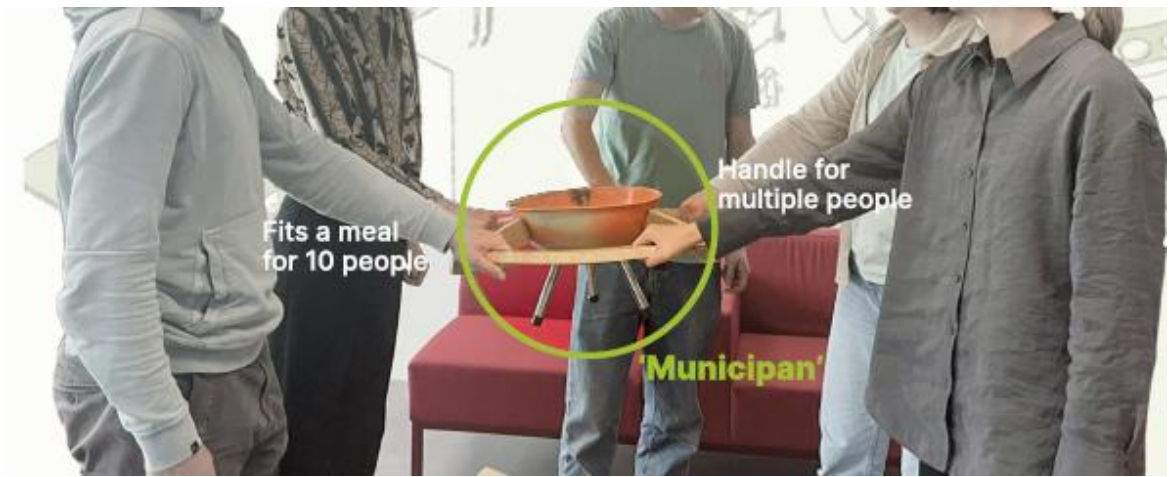

Figure 2: A scale model of Municipan made by the students.

## 3   DESIGNING DE-AUTOMATION

Reflecting on Municipan, it becomes clear that there is no market for such technology in today's societies, as society feeds prosperity and wellbeing based on growth. However, Municipan is not meant to directly function as a bottom-up change towards a postgrowth way of living. Instead, we believe that a bottom-up change can be obtained through the alternative persona that the students engaged with. This persona can form a nucleus that could lead to a different kind of design outcome in future projects by designers. Such a persona is willing and able to invest time and effort and learn new skills invites designs that work towards *de-automation;* i.e., delegation from devices back to people.

In a concrete sense, this could for example be a navigation system that gradually decreases the amount of routing assistance it offers. Imagine that the system shows the destination like a lighthouse, but leaves the driver with the freedom to find the route themselves. Drivers thus bring in their own knowledge and thus strengthen their sense of competence and self-efficacy. Or imagine other examples, such as a robot vacuum cleaner that only works when the sun is shining (i.e., using solar energy), or a heating system (which we will bring to the workshop as an [example](#)) that slowly lowers the indoor temperature, widening the comfort zone of people (Kuijer & de Koning, 2024).

All of the examples are built on the assumption (i.e., persona) that people are willing to learn and invest time. That may feel esoteric or antiquated. However, here we tend to think that the focus on efficiency and convenience, in the design of technology, is so deeply internalized in society and design that it is initially difficult to imagine other conceptions. Design students who have used a different persona in their studies might incorporate post-growth concepts in their future designs more likely and easily. From a postgrowth perspective, designing for this alternative persona (i.e., user) could decrease technological dependence and thereby has potential to increase autonomy and reduce the resource intensity of everyday life.

## ACKNOWLEDGEMENTS

We want to thank Olivier Blom, Tongbin Qi, Merel van Lieshout, Annet Remijnse, and Simon Nieuweboer for their valuable creative input through designing Municipan and their permission for reprint.

**REFERENCES**

Albert Borgmann. (1984). *Technology and the character of contemporary life: a philosophical inquiry*. University of Chicago Press.

Dahlgren, K., Pink, S., Strengers, Y., Nicholls, L., & Sadowski, J. (2021). Personalization and the Smart Home: questioning techno-hedonist imaginaries. *Convergence: The International Journal of Research into New Media Technologies*, *27*(5), 1155–1169. https://doi.org/10.1177/13548565211036801

Dunbar, R. I. M., & Sosis, R. (2018). Optimising human community sizes. *Evolution and Human Behavior : Official Journal of the Human Behavior and Evolution Society*, *39*(1), 106–111. https://doi.org/10.1016/j.evolhumbehav.2017.11.001

Harry Braverman. (1998). *Labor and Monopoly Capital: The Degradation of Work in the Twentieth Century*. NYU Press.

Kuijer, L., & de Koning, P. (2024). Feeling the Heat: Uncomfortable Design Fictions for Alternative Forms of Summer Comfort. *Proceedings of the Eighteenth International Conference on Tangible, Embedded, and Embodied Interaction*, 1–15. https://doi.org/10.1145/3623509.3633391

Latour, B. (1992). Where are the missing masses? The sociology of a few mundane artifacts. In Wiebe E. Bijker & John Law (Eds.), *Shaping Technology/Building Society: Studies in Sociotechnical Change* (pp. 225–258). MIT Press. http://www.open.edu/openlearn/societywww.open.edu/openlearnLatour,B.

Pirgmaier, E. (2020). Consumption corridors, capitalism and social change. *Sustainability: Science, Practice and Policy*, *16*(1), 274–285. https://doi.org/10.1080/15487733.2020.1829846

Strand, R., Saltelli, A., Giampietro, M., Rommetveit, K., & Funtowicz, S. (2018). New narratives for innovation. *Journal of Cleaner Production*, *197*, 1849–1853. https://doi.org/10.1016/j.jclepro.2016.10.194

Tonkinwise, C. (2018). 'I prefer not to.' In *Undesign* (pp. 74–84). Routledge. https://doi.org/10.4324/9781315526379-7

Verbeek, P.-P. (2011). *Moralizing technology: understanding and designing the morality of things*. University of Chicago Press.
4